\documentclass[11pt,a4paper]{article}

\usepackage{jcappub}
\usepackage{graphicx}
\usepackage{amssymb}
\usepackage{amsmath}
\usepackage{bm}
\usepackage{latexsym}
\usepackage{psfrag}
\usepackage{ulem}
\usepackage{textcomp}
\usepackage{color}
\usepackage{pstricks}
\usepackage{float}
\usepackage{subcaption}

\makeatletter
\gdef\@fpheader{}
\makeatother

\def\benu{\begin{enumerate}}
\def\eenu{\end{enumerate}}
\def\nn{\nonumber} 

\def\f{\frac}
\def\l{\left}
\def\r{\right}
\def\d{{\rm d}}
\def\cR{{\mathcal R}}

\def\ei{\eta_{\rm i}}

\def\ee{\eta_{\rm e}}

\def\vk{{\bm k}}
\def\vka{{\bm k}_1}
\def\vkb{{\bm k}_2}
\def\vkc{{\bm k}_3}
\def\ska{{k_1}}
\def\skb{{k_2}}
\def\skc{{k_3}}

\def\cB{{\cal B}}
\def\cG{{\cal G}}

\def\fnl{f_{_{\rm NL}}}

\def\cnls{C_{_{\rm NL}}^{\mathcal R}}

\def\Mpl{M_{_{\rm Pl}}}
\def\Mp{M_{_{\rm Pl}}}

\newcommand{\sr}{{\cal R}}
\newcommand{\g}{\gamma}                 
                 
\newcommand{\kt}{k_{_{\rm T}}}
\newcommand{\ka}{k_{_{{\rm T}1}}}

\newcommand{\kaa}{k_1}
\newcommand{\kbb}{k_2}
\newcommand{\kcc}{k_3}

\newcommand{\fk}{f_k}
\newcommand{\fka}{f_{k_1}}
\newcommand{\fkb}{f_{k_2}}

\newcommand{\hk}{h_k}
\newcommand{\gk}{g_k}

\newcommand{\gkc}{g_{k_3}}
\newcommand{\ck}{c_k}
\newcommand{\cka}{c_{k_1}}
\newcommand{\ckb}{c_{k_2}}

\newcommand{\dk}{d_k}

\newcommand{\phk}{\phi_k}
\newcommand{\phka}{\phi_{k_1}}

\newcommand{\RNum}[1]{\uppercase\expandafter{\romannumeral #1\relax}}

\newcommand{\bea}{\begin{eqnarray}}
\newcommand{\eea}{\end{eqnarray}}

\def\ps{{\mathcal P}_{_{\rm S}}}
\def\pt{{\mathcal P}_{_{\rm T}}}
\def\ns{n_{_{\rm S}}}
\def\nt{n_{_{\rm T}}}

\newcommand{\viz}{\textit{viz.~}}
\newcommand{\ie}{\textit{i.e.~}}
\begin{document}
%%%%%%%%%%%%%%%%%%%%%%%%%%%%%%%%%%%%%%%%%%%%%%%%%%%%%%%%%%%%%%%%%%%%%%%%%%%%%%%
\title{The scalar-scalar-tensor inflationary three-point function
in the axion monodromy model}
\author[a]{Debika~Chowdhury,}
\affiliation[a]{Department of Physics, Indian Institute of 
Technology Madras, Chennai~600036, India}
\emailAdd{debika@physics.iitm.ac.in}
\author[b]{V.~Sreenath}
\affiliation[b]{Department of Physics 
and Astronomy, Louisiana State University, Baton Rouge, LA~70803,
U.~S.~A.}
\emailAdd{sreenath@lsu.edu}
\author[a]{and L.~Sriramkumar}
\emailAdd{sriram@physics.iitm.ac.in}
\date{today} 
%%%%%%%%%%%%%%%%%%%%%%%%%%%%%%%%%%%%%%%%%%%%%%%%%%%%%%%%%%%%%%%%%%%%%%%%%%%%%%%
\abstract{The axion monodromy model involves a canonical scalar field 
that is governed by a linear potential with superimposed modulations.
The modulations in the potential are responsible for a resonant behavior which
gives rise to persisting oscillations in the scalar and, to a smaller extent, 
in the tensor power spectra.
Interestingly, such spectra have been shown to lead to an improved fit to the 
cosmological data than the more conventional, nearly scale invariant, primordial 
power spectra. 
The scalar bi-spectrum in the model too exhibits continued modulations and the 
resonance is known to boost the amplitude of the scalar non-Gaussianity parameter
to rather large values.
An analytical expression for the scalar bi-spectrum had been arrived at earlier
which, in fact, has been used to compare the model with the cosmic microwave
background anisotropies at the level of three-point functions involving scalars.  
In this work, with future applications in mind, we arrive at a similar analytical 
template for the scalar-scalar-tensor cross-correlation.
We also analytically establish the consistency relation (in the squeezed limit)
for this three-point function.
We conclude with a summary of the main results obtained.}
\maketitle

%%%%%%%%%%%%%%%%%%%%%%%%%%%%%%%%%%%%%%%%%%%%%%%%%%%%%%%%%%%%%%%%%%%%%%%%%%%%%%%

\section{Introduction}\label{sec:introduction}

Until very recently, inflationary models were compared with the data at the level 
of two-point functions, \ie the inflationary scalar and tensor power spectra were 
compared with the angular power spectra of the Cosmic Microwave Background (CMB) 
and the matter power spectrum associated with the Large Scale Structure 
(LSS)~\cite{Mortonson:2010er,Easther:2011yq,Norena:2012rs,Martin:2010hh,
Martin:2013tda,Martin:2013nzq,Martin:2014lra,Martin:2014rqa,Ade:2015lrj}.
Over the last decade and a half, it has been realized that non-Gaussianities in 
general and the three-point functions in particular can provide strong constraints 
on the physics of the early universe.
On one hand, there has been tremendous progress in understanding the generation 
of non-Gaussianities during inflation~\cite{Gangui:1993tt,Gangui:1994yr,Gangui:1999vg,
Wang:1999vf,Maldacena:2002vr,Seery:2005wm,Chen:2005fe,Chen:2006nt,Langlois:2008wt,
Langlois:2008qf,Chen:2010xka,Wang:2013zva,Martin:2015dha} and the corresponding 
signatures on the CMB and the LSS~\cite{Komatsu:2001rj,Komatsu:2003iq,Babich:2004yc,
Liguori:2005rj,Hikage:2006fe,Fergusson:2006pr,Yadav:2007rk,Creminelli:2006gc,
Yadav:2007yy,Hikage:2008gy,Rudjord:2009mh,Smith:2009jr,Smidt:2010ra,Fergusson:2010dm,
Liguori:2010hx,Yadav:2010fz,Komatsu:2010hc}. 
On the other hand, the expectation alluded to above has been largely corroborated
by the strong constraints that have been arrived at from the Planck CMB data on the 
scalar non-Gaussianity parameter $\fnl$~\cite{Ade:2015ava}.
The recent observations seem to suggest that the primordial perturbations are 
consistent with a Gaussian distribution.

\par

A nearly scale invariant primordial scalar power spectrum is remarkably consistent 
with the observations of the CMB~\cite{Mortonson:2010er,Easther:2011yq,Norena:2012rs,
Martin:2013tda, Martin:2013nzq, Martin:2014lra,Martin:2010hh,Martin:2014rqa,Ade:2015lrj}.
However, it has been repeatedly noticed that certain features in the inflationary power 
spectrum can improve the fit to the data (for instance, see Refs.~\cite{Jain:2008dw,
Jain:2009pm,Hazra:2010ve,Hazra:2013xva,Hazra:2014jwa,Hunt:2013bha,Hunt:2015iua,Hazra:2014jka,
Hazra:2014goa}).
One such type of feature is continued oscillations in the scalar power spectrum that 
extends over a wide range of scales~\cite{Martin:2003sg,Martin:2004iv,Martin:2004yi,
Zarei:2008nr,Pahud:2008ae,Kobayashi:2010pz,Meerburg:2013cla,Meerburg:2013dla}. 
Such a power spectrum is known to be generated by the so-called axion monodromy model, 
motivated by string theory~\cite{Flauger:2009ab,Aich:2011qv,Peiris:2013opa, 
Easther:2013kla,Meerburg:2014bpa}.
The model is described by a linear potential with superimposed oscillations. 
The oscillations in the potential give rise to a resonant behavior which leads to 
continued modulations in the scalar and tensor power spectra. 
At the cost of two or three extra parameters, the resulting power spectra are 
known to improve the fit to the CMB data from the Wilkinson Microwave Anisotropy Probe 
(WMAP) and Planck by as much as $\Delta \chi^2\simeq 10$--$20$~\cite{Aich:2011qv,
Peiris:2013opa,Easther:2013kla,Meerburg:2013cla,Meerburg:2013dla}. 
Ideally, one would like to carry out a similar analysis of comparing the model with
the CMB data at the level of the three-point functions as well. 
But, it proves to be numerically taxing to evaluate the three-point functions in 
these  models (see, for instance, Ref.~\cite{Hazra:2012yn}). 
In such a situation, clearly, it will be convenient if there exist analytical 
templates for the inflationary three-point functions. 
Such a template for the scalar bi-spectrum has been arrived at 
earlier~\cite{Flauger:2009ab,Flauger:2010ja}, which has been utilized towards 
comparing models leading to oscillatory features with the CMB 
data~\cite{Fergusson:2014hya,Fergusson:2014tza}.

\par

Apart from the scalar bi-spectrum, there exist three other three-point functions
which involve the tensor perturbations~\cite{Maldacena:2002vr}.
Amongst these three-point functions, the scalar-scalar-tensor cross-correlation 
has the largest amplitude after the scalar bi-spectrum (in this context, see
Refs.~\cite{Maldacena:2002vr,Gao:2012ib,Dai:2013ikl,Sreenath:2013xra,Kundu:2013gha,
Dimastrogiovanni:2014ina,Dimastrogiovanni:2015pla,Maldacena:2011nz,Gao:2011vs}).    
In this work, motivated by the efforts towards arriving at an analytical template
for the scalar bi-spectrum in the axion monodromy model, we obtain a similar
template for the scalar-scalar-tensor three-point function for an arbitrary 
triangular configuration of the wavevectors.
In the case of the scalar bi-spectrum, in order to determine the dominant contribution
due to the oscillations in the potential, it was sufficient to take into account the 
effects due to the changes in the behavior of the slow roll parameters, and one could 
work with the simple de Sitter modes for the curvature perturbation.
In contrast, to evaluate the scalar-scalar-tensor cross correlation, we find that
apart from the changes in the behavior of the slow roll parameters, we also need 
to take into account the modifications to the de Sitter modes.
As in the purely scalar case~\cite{Creminelli:2004yq,Cheung:2007sv,RenauxPetel:2010ty,
Ganc:2010ff,Creminelli:2011rh,Chialva:2011iz,Schalm:2012pi,Pajer:2013ana,Kenton:2015lxa,
Kenton:2016abp}, the other 
three-point functions involving the tensors are also known to satisfy the so-called 
consistency relation in the squeezed limit~\cite{Dai:2013ikl,Kundu:2013gha,
Sreenath:2014nka,Dimastrogiovanni:2014ina,Kundu:2014gxa,Kundu:2015xta}.
We shall analytically establish the consistency condition for the scalar-scalar-tensor 
three-point function in the axion monodromy model.

\par

The remainder of this paper is organized as follows. 
In the next section, we shall briefly discuss the important aspects of the
axion monodromy model and arrive at the scalar and tensor power spectra in
the model.
In Sec.~\ref{sec:amm-itpfs}, we shall gather the essential expressions describing
the scalar-scalar-tensor three-point function and the corresponding non-Gaussianity 
parameter.
In Sec.~\ref{sec:amm-sstcc}, we shall arrive at an analytical expression for the 
scalar-scalar-tensor cross-correlation under certain approximations.
To illustrate the extent of accuracy of the approximations, we shall also compare 
the analytical result with the corresponding numerical result.
Further, in Sec.~\ref{sec:sqlcr}, we shall analytically verify the consistency 
relation for the three-point function in the squeezed limit.
We shall conclude in Sec.~\ref{sec:d} with a brief summary of the results obtained.

\par

A few remarks on our conventions and notations seem essential at this stage
of our discussion.
We shall work with natural units wherein $\hbar=c=1$, and define the Planck 
mass to be $\Mpl=(8\,\pi\, G)^{-1/2}$. 
We shall adopt the signature of the metric to be $(-,+,+,+)$. 
We shall assume the background to be the spatially flat 
Friedmann-Lema\^itre-Robertson-Walker (FLRW) line element that is described by 
the scale factor $a$ and the Hubble parameter $H$. 
As is convenient, we shall switch between various parameterizations of time, 
\viz the cosmic time~$t$, the conformal time~$\eta$ or e-folds denoted by~$N$. 
An overdot and an overprime shall represent differentiation with respect to the
cosmic and the conformal time coordinates, respectively.
We shall restrict our attention in this work to inflationary models involving 
the canonical scalar field. 
Note that, in such a case, the first and second slow roll parameters are defined 
as $\epsilon_1 = -\dot H / H^2$ and $\epsilon_2 = \d \ln \,\epsilon_1/\d N$.  

%%%%%%%%%%%%%%%%%%%%%%%%%%%%%%%%%%%%%%%%%%%%%%%%%%%%%%%%%%%%%%%%%%%%%%%%%%%%%%%

\section{The axion monodromy model}\label{sec:amm}

In this section, we shall summarize the essential aspects of the 
axion monodromy model.
We shall discuss the evolution of the background as well as the evolution of 
the scalar and tensor perturbations and also arrive at the corresponding power 
spectra. 

%%%%%%%%%%%%%%%%%%%%%%%%%%%%%%%%%%%%%%%%%%%%%%%%%%%%%%%%%%%%%%%%%%%%%%%%%%%%%%%

\subsection{The evolution of the background and the slow roll parameters}

The axion monodromy model is described by a linear potential with 
superimposed oscillations~\cite{Aich:2011qv,Flauger:2009ab,Peiris:2013opa,
Easther:2013kla}.
The potential is given by
\begin{equation}
V(\phi)=\mu^3\, \phi+\mu^3\, b\, f\, {\rm cos}\,\l(\f{\phi}{f}\r),
\label{eq:amm}
\end{equation}
where $b$ is a dimensionless quantity and we have ignored a possible phase 
factor in the trigonometric function.
Throughout this work, we shall assume that $b$ is small and attempt to derive 
all the results at the linear order in $b$ (as we shall discuss in due course, 
the constraint from the recent Planck data suggests that $b$ is indeed small,
of the order of $10^{-2}$).
Evidently, this assumption is equivalent to considering the trigonometric 
modulations as small departures from the linear potential. 
We shall also assume that the linear potential admits slow roll and that the 
modulations lead to deviations from the monotonic slow roll 
behavior~\cite{Flauger:2009ab}.

\par

The equation of motion governing the scalar field described by the 
potential~(\ref{eq:amm}) above is given by
\begin{equation}
{\ddot \phi}+3\, H\, {\dot \phi}
+\mu^3-\mu^3\, b\, {\rm sin}\l(\f{\phi}{f}\r)=0,\label{eq:de-sf} 
\end{equation}
with the Hubble parameter $H = {\dot a}/a$ being determined by the first Friedmann equation, 
\viz\/
\begin{equation}
H^2=\f{1}{3\,\Mpl^2}\l[\f{\dot \phi^2}{2} + V(\phi)\r].\label{eq:ffe}
\end{equation}
Since we shall assume that $b$ is small, we can write the background inflaton 
as a slowly rolling part plus a part which describes the modulations as 
\begin{equation}
\phi=\phi_0+b\, \phi_1+\ldots. 
\end{equation}
As we mentioned, we shall limit ourselves to terms which are linear in $b$. 
At the leading order, under the slow roll approximation, the 
equations~(\ref{eq:de-sf}) and~(\ref{eq:ffe}) simplify to
\begin{subequations}
\bea
3\, H\, {\dot \phi}_0 &\simeq& -\mu^3,\\
3\, H^2\, \Mpl^2 &\simeq& \mu^3\, \phi_0.
\eea
\end{subequations}
These equations can be easily integrated to yield the leading order term in 
the inflaton to be
\begin{equation}
\phi_0(t)=\l[\phi_{\ast}^{3/2}
-\f{\sqrt{3\,\mu^3}}{2}\,\Mpl\, \l(t-t_\ast\r)\r]^{2/3},\label{eq:s-phi0}
\end{equation}
where $\phi_0(t_\ast)=\phi_\ast$, with $t_\ast$ denoting the time when the 
pivot scale, say, $k_\ast$, leaves the Hubble radius.
It should be noted that, in order to achieve about $60$--$70$ e-folds
of inflation, we shall require that $\phi_\ast\simeq 10\,\Mpl$.

\par

Let us now consider the behavior of $\phi_1$.
The differential equation satisfied by the component $\phi_1$ can be obtained 
to be
\begin{equation}
{\ddot \phi}_1+\f{\sqrt{3\,\mu^3\, \phi_0}}{\Mpl}\, {\dot \phi}_1
-\f{\mu^3}{2\,\phi_0}\, \phi_1=\mu^3\, {\rm sin}\l(\f{\phi_0}{f}\r),
\end{equation}
where we have made use of the fact that, until the first order in $b$, under 
the slow roll approximation, we can write 
\begin{equation}
H^2\simeq\f{\mu^3}{3\, \Mpl^2}\, \l(\phi_0+ b\,\phi_1\r). 
\end{equation}
If we make use of the solution~(\ref{eq:s-phi0}) for $\phi_0$, the above 
equation governing $\phi_1$ can be expressed as
\begin{equation}
\f{\d^2\phi_1}{\d\phi_0^2}-\f{3\,\phi_0}{\Mpl^2}\,\f{\d\phi_1}{\d\phi_0}
-\f{3\,\phi_1}{2\,\Mpl^2}=\f{3\,\phi_0}{\Mpl^2}\, {\rm sin} \l(\f{\phi_0}{f}\r). 
\end{equation}
This equation can be integrated to arrive at the following solution for
$\phi_1$:
\begin{equation}
\phi_1(\phi_0)
=\f{3\,f^2\,\phi_\ast/\Mpl^2}{1+(3\,f\,\phi_\ast/\Mpl^2)^2}\,
\l[-{\rm sin}\l(\f{\phi_0}{f}\r)+\f{3\,f\,\phi_\ast}{\Mpl^2}\,
{\rm cos}\l(\f{\phi_0}{f}\r)\r],
\end{equation}
which can be written as
\begin{equation}
\phi_1(\phi_0) 
=-\f{3\,f^2\,\phi_\ast/\Mpl^2}{\l[1+(3\,f\,\phi_\ast/\Mpl^2)^2\r]^{1/2}}\,
{\rm sin}\l(\f{\phi_0}{f}-\psi_1\r),
\end{equation}
where 
\begin{equation}
{\rm sin}\,\psi_1
=\f{3\,f\,\phi_\ast/\Mpl^2}{\l[1+(3\,f\,\phi_\ast/\Mpl^2)^2\r]^{1/2}}\;. 
\end{equation}
Note that, when $f\,\phi_\ast/\Mpl^2$ is assumed to be small (as we shall 
discuss later, the constraints from the recent CMB observations suggest
that $f\,\phi_\ast/\Mpl^2\simeq 7.6346\times 10^{-2}$), we can write
\begin{equation}
\phi_1(\phi_0) 
=-\f{3\,f^2\,\phi_\ast}{\Mpl^2}\,
{\rm sin}\l(\f{\phi_0}{f}\r).
\end{equation}

\par

Having understood the behavior of the background inflaton, let us now evaluate 
the first and the second slow roll parameters. 
Recall that the first slow roll parameter is given by $\epsilon_1=-{\dot H}/H^2
={\dot \phi}^2/(2\, H^2\,\Mpl^2)$.
Let us write the first slow roll parameter as $\epsilon_1=\epsilon_1^0
+\epsilon_1^{\rm c}$, where $\epsilon_1^0$ is the contribution due to
$\phi_0$, while $\epsilon_1^{\rm c}$ is the correction at the first
order in $b$.
As one would expect, $\epsilon_1^0$ will roughly be constant and will 
be of the order of $\epsilon_1^0=\epsilon_1^\ast\simeq 
\Mpl^2/(2\,\phi_\ast^2)$, determined by the linear term in the potential.
Upon making use of the solutions we have obtained above, we can show 
that, the quantity $\epsilon_1^{\rm c}$ is given by
\begin{equation}\label{eq:eps1c}
\epsilon_1^{\rm c}
= -\f{3\,b\,f}{\phi_\ast}
\cos\l(\f{\phi_0}{f}\r).
\end{equation}
In arriving at this expression, we have again made the assumption that $f\,
\phi_\ast/\Mpl^2 \ll 1$.
The second slow roll parameter can be expressed as $\epsilon_2
={\dot \epsilon}_1/(H\, \epsilon_1)$.
One can show that $\epsilon_2=2\, (\delta+\epsilon_1)$, where $\delta 
={\ddot H}/(H\, {\dot H})={\ddot \phi}/(H\, {\dot \phi})$. 
If we write $\delta=\delta_0+\delta_1$, where $\delta_0$ corresponds to the 
case wherein $b$ vanishes, then we can show that $\delta_0=\epsilon_1^{\ast}$ 
and $\epsilon_2^*=4\,\epsilon_1^\ast$.
Also, for small $f\,\phi_\ast/\Mpl^2$, we find that $\delta_1$ can be 
expressed as
\begin{equation}\label{eq:delta1}
\delta_1 = -3\,b\,\sin\l(\f{\phi_0}{f}\r). 
\end{equation}

%%%%%%%%%%%%%%%%%%%%%%%%%%%%%%%%%%%%%%%%%%%%%%%%%%%%%%%%%%%%%%%%%%%%%%%%%%%%%%%

\subsection{The evolution of the perturbations and the power spectra}

The assumptions and approximations that we have made in the previous 
sub-section enable us to analytically explore the evolution of 
perturbations and calculation of the power spectra~\cite{Flauger:2009ab}.

\par

Let us begin by summarizing a few basic points concerning the scalar power 
spectrum.
Recall that, upon quantization, the curvature perturbation $\cR$ can be 
decomposed in terms of the corresponding Fourier modes as follows:
\begin{equation}
\hat{\cR}(\eta, {\bf x}) 
= \int \frac{\d^{3}{\bm k}}{\l(2\,\pi\r)^{3/2}}\,
\hat{\cR}_{\bm k}(\eta)\, {\rm e}^{i\,{\bm k}\cdot{\bm x}}
= \int \frac{\d^{3}{\bm k}}{\l(2\,\pi\r)^{3/2}}\,
\l(\hat{a}_{\bm k}\,f_{k}(\eta)\, 
{\rm e}^{i\,{\bm k}\cdot{\bm x}}
+\hat{a}^{\dagger}_{\bf k}\,f^{*}_{k}(\eta)\,
{\rm e}^{-i\,{\bm k}\cdot{\bm x}}\r),\label{eq:s-m-dc}\\
\end{equation}
where the modes $f_k$ satisfy the differential equation
\begin{equation}
f_k''+2\,\frac{z'}{z}\,f_k'+k^2\,f_k = 0,\label{eq:de-fk}\\
\end{equation}
with $z=\sqrt{2\,\epsilon_1}\,a\,\Mpl$, while the annihilation and the 
creation operators ${\hat a}_{\bm k}$ and $\hat{a}^{\dagger}_{\bm k}$ 
obey the standard commutation relations.
The scalar power spectrum, \viz\/ ${\mathcal P}_{_{\rm S}}(k)$, is defined 
as 
\begin{equation}
\langle\, {\hat \cR}_{\bm k}(\eta)\,
{\hat \cR}_{\bm k'}(\eta)\,\rangle
=\f{(2\,\pi)^2}{2\, k^3}\, {\mathcal P}_{_{\rm S}}(k)\;
\delta^{(3)}({\bm k}+{\bm k'}),\label{eq:sps-d}\\
\end{equation}
where the expectation value on the left hand side is to be evaluated in 
the specified initial quantum state of the perturbations.
If one assumes the initial state of the perturbations to be the vacuum state 
$\vert 0\rangle$ (defined as ${\hat a}_{\bm k}\vert 0\rangle=0$ $\forall\,\vk$) 
then, on making use of the decomposition~(\ref{eq:s-m-dc}) in the above 
definition, the inflationary scalar power spectrum ${\mathcal P}_{_{\rm S}}(k)$ 
can be expressed as 
\begin{equation}
{\mathcal P}_{_{\rm S}}(k)
=\f{k^3}{2\, \pi^2}\, \vert f_k\vert^2.\label{eq:sps}
\end{equation}
The amplitude $\vert f_k\vert$ on the right hand side of the above expression 
is to be evaluated when the modes are sufficiently outside the Hubble radius.
It is useful to note here that the scalar spectral index $\ns$ is defined as 
\begin{equation}
\ns=1+\f{\d \ln \ps(k)}{\d \ln k}.\label{eq:ns}\\ 
\end{equation}

\par

Let us analytically evaluate the scalar power spectrum in the axion 
monodromy model~\cite{Flauger:2009ab}.
If we choose to work in terms of the new variable $x = -k\,\eta$ and 
use the exact relation 
\begin{equation}
\f{z'}{z } = a\,H\,\l( 1 + \epsilon_1+\delta \r),
\end{equation}
we can rewrite the differential equation~(\ref{eq:de-fk}) governing 
the mode $f_k$ as
\begin{equation}
\f{\d^2f_k}{\d x^2}
+2\,\l(1+\epsilon_1+\delta\r)\, \l(\f{a\, H}{-k}\r)\,\f{\d f_k}{\d x}
+f_k=0.
\end{equation}
At the leading order in slow roll, we have 
\begin{equation}
a\,H\simeq -\f{1+\epsilon_1}{\eta}.\label{eq:aH}
\end{equation}
We shall work in the de Sitter approximation when $b=0$, which corresponds
to ignoring the contributions due to the $\epsilon_1^\ast$ term.
The effects due to the $\delta_1$ term dominate the effects due to the
$\epsilon_1^{\rm c}$ term.
Under these conditions, we can write the above differential equation 
describing $f_k$ as follows:
\begin{equation}
\f{\d^2 f_k}{\d x^2}-\f{2\,\l(1+\delta_1\r)}{x}\,\f{\d f_k}{\d x}
+f_k=0.\label{eq:de-cRk}
\end{equation}

\par

In the slow roll limit determined by the linear potential wherein $\delta_1$ 
can be ignored, the positive frequency modes satisfying the above differential 
equation can be written in the de Sitter form as
\begin{equation}
\fk^+(x)=i\, \fk^0\, \l(1-i\,x\r)\, {\rm e}^{i\,x}, 
\end{equation}
where $\fk^0=H_0/(2\, \Mpl\,\sqrt{k^3\,\epsilon_1^\ast})$, 
with $H_0$ being given by $H_0^2=\mu^3\,\phi_\ast/(3\,\Mp^2)$.
Hence, in the presence of a non-zero $\delta_1$, let us write the modes 
describing the curvature perturbation as~\cite{Flauger:2009ab}
\begin{equation}
\fk(x) = \fk^+(x)+\ck(x)\, \fk^-(x),\label{eq:amm-s-f}
\end{equation}
where $\fk^-(x)$ are the negative frequency modes that are related to the 
positive frequency modes by the relation $\fk^-(x)=\fk^{+\ast}(x)$.
Note that the non-vanishing $\delta_1$ modifies the standard 
de Sitter modes $\fk^+(x)$. The non-trivial evolution of the modes is 
captured by the function $\ck(x)$.

The de Sitter modes $\fk^{\pm}$ satisfy the differential equation
\begin{equation}
\f{\d^2\fk^{\pm}}{\d x^2}
-\f{2}{x}\, \f{\d \fk^{\pm}}{\d x}+\fk^{\pm}=0.\label{eq:de-dSm} 
\end{equation}
Therefore, upon substituting the expression~(\ref{eq:amm-s-f}) in the
differential equation~(\ref{eq:de-cRk}), we obtain the following
equation governing $c_k$:
\begin{equation}\label{eq:ck-exact}
\f{\d}{\d x}\l[\l(1-\f{i}{x}\r)\, {\rm e}^{-2\,i\,x}\, 
\f{\d \ck}{\d x}\r]
+\f{i}{x^2}\, {\rm e}^{-2\,i\,x}\, \f{\d \ck}{\d x}
=\f{2\,i\,\delta_1}{x}.
\end{equation}
As is well known, the perturbations oscillate when they are well inside
the Hubble radius.
In this sub-Hubble regime, the oscillations in the perturbations resonate
with the oscillations in the background quantities.
This resonance occurs when $x\simeq
\Mpl^2/(2\,f\,\phi_\ast)$, which proves to be a large quantity for the 
parameter ranges of our interest (recall that, we had assumed $f\,
\phi_\ast/\Mpl^2$ to be small).
Therefore, the terms involving inverse powers of $x$ on the left hand side
of the above differential equation can be ignored in the sub-Hubble regime 
and, under these conditions, the equation can be easily integrated.
We find that the resulting $\ck(x)$ can be expressed as
\begin{eqnarray}
\ck(x)
&=&-\f{3\,b\,f\,\phi_\ast}{2\,\Mpl^2}\,
\biggl[{\rm e}^{i\,(\alpha_1 + \phk/f)}\,
{\rm e}^{-\pi\,\Mpl^2/(2\,f\,\phi_\ast)}\;
\Gamma\l(1+\f{i\,\Mpl^2}{f\,\phi_\ast},-2\,i\,x\r)\nn\\
& &+\,{\rm e}^{-i\,(\alpha_1 + \phk/f)}\,
{\rm e}^{\pi\,\Mpl^2/(2\,f\,\phi_\ast)}\;
\Gamma\l(1-\f{i\,\Mpl^2}{f\,\phi_\ast},-2\,i\,x\r)\biggr],
\label{eq:ck-full}
\end{eqnarray}
where $\Gamma(a,x)$ is the incomplete Gamma 
function~\cite{Gradshteyn:2007} and $\alpha_1 = -X_{\rm res}\,{\rm ln}\,2$,
with $X_{\rm res} = \Mp^2/(f\,\phi_\ast)$. Note that $X_{\rm res}$ is a large
quantity since we have assumed $f\,\phi_\ast/\Mp^2$ to be small.
Note that, in arriving at the above expression, we have expressed $\phi_0$ 
in terms of $x$ as
\begin{equation}
\phi_0 = \phk\,+\,\sqrt{2\,\epsilon_1^*}\,\Mp\,{\rm ln}\,x,
\label{eq:amm-phi0}
\end{equation}
where the quantity $\phk$ is given by
\begin{equation}
\phk=\phi_\ast-\sqrt{2\,\epsilon_1^{\ast}}\, \Mpl\,
{\rm ln}\l(\f{k}{k_\ast}\r)\label{eq:phik} 
\end{equation}
with $k_\ast$ denoting the pivot scale.
Since we have assumed that $f\,\phi_\ast/\Mp^2 \ll 1$, the first term in
Eq.~(\ref{eq:ck-full}) is exponentially suppressed and hence can be ignored. 
Thus, we can express $\ck$ as 
\begin{equation}
\ck(x)= 
-\f{3\,b\,f\,\phi_\ast}{2\,\Mpl^2}\,
{\rm e}^{-i\,\l(\phk/f+\alpha_1\r)}\,{\rm e}^{\pi\,\Mp^2/2\,f\,\phi_\ast\,}\,
\Gamma\l(1-\f{i\,\Mpl^2}{f\,\phi_\ast},-2\,i\,x\r).\label{eq:ck}
\end{equation}

Note that, in deriving the above solution for $\ck(x)$, we had 
ignored inverse powers of $x$ on the left hand side of Eq.~(\ref{eq:ck-exact}).
We should emphasize here that this approximation is strictly valid only at 
sub-Hubble scales. However, since the complete mode approaches a constant value 
at late times, one finds that the largest contribution to the three-point functions 
arises from the sub-Hubble domain~\cite{Flauger:2009ab}. Hence, the above solution 
for $\ck(x)$ proves to be sufficient for evaluating the three-point function of our
interest analytically. As we shall see, these arguments are corroborated by the
numerical results we obtain.

\par

The scalar power spectrum can now be obtained from the late time limit (\ie\/
as $x\to 0$) of the modes $f_k$.
We find that 
\begin{equation}\label{eq:ck0}
\ck(0) = \frac{3\,i\,b\,\sqrt{\pi}}{\sqrt{2\,X_{\rm res}}}\,
{\rm e}^{-i\,\l(\phk/f+\beta_1\r)},
\end{equation}
where the phase $\beta_1 = X_{\rm res}\,{\rm ln}\,X_{\rm res} - X_{\rm res} 
- X_{\rm res}\,{\rm ln}\,2 - \pi/4$.
Upon using this result, at the order $b$, 
we can express the scalar power spectrum as~\cite{Flauger:2009ab}
\begin{equation}
\ps(k)
=\ps^0\, \l[1-\ck(0)-\ck^\ast(0)\r]
= \ps^0\,\l[1-\delta\ns\,\sin\l(\f{\phk}{f}+\beta_1\r)\r],\label{eq:amm-ps}
\end{equation}
where $\ps^0$ represents the amplitude of the scalar power spectrum which arises 
in the slow roll scenario when the oscillations in the potential are absent
and the quantity $\phi_k$ depends on the wavenumber through the 
relation~(\ref{eq:phik}). 
The quantity $\ps^0$ is given by
\begin{equation}
\ps^0 = \frac{H_0^2}{8\,\pi^2\,\Mp^2\,\epsilon_1^\ast}\label{eq:amm-ps0}
\end{equation}
and, for small $f\,\phi_\ast/\Mpl^2$, the quantity $\delta\ns$ can be expressed 
as 
\begin{equation}
\delta\ns 
=\f{3\,b\,\sqrt{2\pi}}{\sqrt{X_{\rm res}}}.\label{eq:dns}
\end{equation}
The sinusoidal term in the power spectrum leads to oscillations that extends over 
a wide range of scales.
These oscillations result in continued modulations in the scalar spectral index 
$\ns$ [cf.~Eq~.~(\ref{eq:ns})], which can be obtained from the 
expression~(\ref{eq:amm-ps}) for the scalar power spectrum.
In order to separate the contributions at the zeroth and first order in $b$, it is 
convenient to write the scalar spectral index in the form $\ns =\ns^0+\ns^{\rm c}$, 
where $\ns^0$ is the scalar spectral index in the slow roll approximation when the 
oscillations are ignored and $\ns^{\rm c}$ is the correction at order $b$. 
One can show that $\ns^0 = 1- 2\,\epsilon_1^\ast - \epsilon_2^\ast
=1-6\,\epsilon_1^\ast$.
The first order correction to scalar spectral index can be evaluated 
from Eqs.~(\ref{eq:ns}) and~(\ref{eq:amm-ps}) to be 
\begin{equation}
\ns^{\rm c}(k) 
= 3\,b\,\sqrt{2\,\pi\,X_{\rm res}}\,
\cos \l(\f{\phk}{f}+\beta_1\r).\label{eq:amm-nsc}
\end{equation}
The power spectrum~(\ref{eq:amm-ps}) that we have arrived at above has been 
compared with the WMAP and Planck data~\cite{Easther:2013kla,Meerburg:2013cla,
Meerburg:2013dla}.
As we have discussed earlier, the persistent oscillations in the power spectrum
lead to a better fit to the data than the more conventional nearly scale
invariant primordial spectrum.
The values of the parameters describing the axion monodromy model that are found 
to lead to the best fit to the Planck data are as follows: 
$\mu/\Mpl=2.512 \times 10^{-10}$, 
$b=1.063\times10^{-2}$ and $f/\Mpl=7.6346\times 10^{-3}$~\cite{Meerburg:2013dla}.
Note that these values lead to $f\, \phi_\ast/\Mpl^2= 7.6346\times 10^{-2}$,
which in turn corresponds to $X_{\rm res}\simeq 13$.
It should be mentioned that earlier CMB data had suggested values for $f$ that
was smaller by an order of magnitude or more and hence a suitably larger value 
of $X_{\rm res}$ (of the order of $250$ or so).
While $X_{\rm res}\simeq 13$ is not very large, we find that our analytical 
results match the numerical results fairly well over a range of $f$ and $b$.

\par

Let us now turn to the case of the tensor power spectrum.
On quantization, the tensor perturbation $\gamma_{ij}$ can be decomposed 
in terms of the corresponding Fourier modes as
\bea 
\hat{\gamma}_{ij}(\eta, {\bf x}) 
&=& \int \frac{\d^{3}{\bm k}}{\l(2\,\pi\r)^{3/2}}\,
\hat{\gamma}_{ij}^{\bm k}(\eta)\, {\rm e}^{i\,{\bm k}\cdot{\bm x}}\nn\\
&=& \sum_{s}\int \frac{\d^{3}{\bm k}}{(2\,\pi)^{3/2}}\,
\l(\hat{b}^{s}_{\bm k}\, \varepsilon^{s}_{ij}({\bm k})\,
h_{k}(\eta)\, {\rm e}^{i\,{\bm k}\cdot{\bm x}}
+\hat{b}^{s\dagger}_{\bf k}\,\varepsilon^{s*}_{ij}({\bm k})\, h^{*}_{k}(\eta)\,
{\rm e}^{-i\,{\bm k}\cdot{\bm x}}\r),\label{eq:t-m-dc}
\eea
where the modes $h_k$ satisfy the differential equation
\begin{equation}
\hk''+2\,\frac{a'}{a}\,\hk'+k^2\,\hk = 0.\label{eq:de-hk}
\end{equation}
The quantity $\varepsilon^{s}_{ij}({\bm k})$ represents the polarization tensor
of the gravitational waves, with the index~$s$ denoting the helicity of the 
graviton.
The transverse and traceless nature of the gravitational waves implies that the 
polarization tensor obeys the relations: $\varepsilon^{s}_{ii}({\bm k})=k_{i}\,
\varepsilon_{ij}^s({\bm k})=0$.
We shall choose to work with the normalization $\varepsilon_{ij}^{r}({\bm k})\,
\varepsilon_{ij}^{s}({\bm k})=2\,\delta^{rs}$.
As in the case of scalars, the annihilation and creation operators 
$\hat{b}_{\bm k}^{s}$ and $\hat{b}^{s\dagger}_{\bm k}$ satisfy the conventional 
commutation relations.
The tensor power spectrum, \viz\/ ${\mathcal P}_{_{\rm T}}(k)$, is defined as 
follows:
\begin{eqnarray}
\langle\, {\hat \gamma}_{ij}^{\bm k}(\eta)\,
{\hat \gamma}_{mn}^{\bm k'}(\eta)\,\rangle
&=&\f{(2\,\pi)^2}{2\, k^3}\, \f{\Pi_{ij,mn}^{{\bm k}}}{4}\;
{\mathcal P}_{_{\rm T}}(k)\;
\delta^{(3)}({\bm k}+{\bm k'}),\label{eq:tps-d}
\end{eqnarray}
with the expectation values on the left hand side to be evaluated in the specified 
initial quantum state, and the quantity $\Pi_{ij,mn}^{\vk}$ is given by
\begin{equation}
\Pi_{ij,mn}^{\vk}
=\sum_{s}\;\varepsilon_{ij}^{s}(\vk)\;
\varepsilon_{mn}^{s\ast}(\vk).
\end{equation}
On making use of the decomposition~(\ref{eq:t-m-dc}), the tensor power spectrum
evaluated in the vacuum state $\vert 0\rangle$ (such that $\hat{b}_{\bm k}^{s}
\vert 0\rangle =0$ $\forall\;{\bm k}$ and $s$) can be expressed as 
\begin{equation}
{\mathcal P}_{_{\rm T}}(k)
= 4\;\f{k^3}{2\, \pi^2}\, \vert h_k\vert^2,\label{eq:tps}
\end{equation}
with the right hand side to be evaluated at sufficiently late times when the modes
are well outside the Hubble radius.
The tensor spectral index $\nt$ is defined as 
\begin{equation}
\nt = \f{\d \ln \pt(k)}{\d \ln k}.\label{eq:nt} 
\end{equation}

\par

In the axion monodromy model, the tensor modes and the tensor power spectrum can 
be determined in a manner very similar to the scalar case.
On substituting the expression~(\ref{eq:aH}) in the equation~(\ref{eq:de-hk}) 
governing the evolution of the tensor modes, we obtain that 
\begin{equation}
\f{\d^2 \hk}{\d x^2}
-\f{2\,\l(1+\epsilon_1\r)}{x}\f{\d \hk}{\d x}+\hk=0.\label{eq:de-hk-x}
\end{equation}
When the modulations in the potential are ignored, the positive frequency tensor 
modes in the slow roll limit are given by
\begin{equation}
h_k^+(x)=i\, h_k^0\, \l(1-i\,x\r)\, {\rm e}^{i\,x}, 
\end{equation}
where $h_k^0=H_0/(\Mpl\,\sqrt{k^3})$.
In the presence of the modulations, let us write the modes describing the tensor 
perturbation as 
\begin{equation}
h_k(x) = h_k^+(x)+\dk(x)\, h_k^-(x),\label{eq:amm-s-g}
\end{equation}
where $h_k^-(x)=h_k^{+\ast}(x)$.
The de Sitter modes $h_k^{\pm}$ satisfy the differential 
equation~(\ref{eq:de-dSm}).
On substituting the expression~(\ref{eq:amm-s-g}) for the tensor modes in 
Eq.~(\ref{eq:de-hk}), we find that the quantity $d_k$ satisfies the 
differential equation
\begin{equation}
\f{\d}{\d x}\l[\l(1-\f{i}{x}\r)\, {\rm e}^{-2\,i\,x}\, 
\f{\d \dk}{\d x}\r]
+\f{i}{x^2}\, {\rm e}^{-2\,i\,x}\, \f{\d \dk}{\d x}
=\f{2\,i\,\epsilon_1}{x}.
\end{equation}
As in the scalar case, due to the resonance that arises in the sub-Hubble
regime (\ie\/ when $x$ is large), 
we can ignore the terms involving the inverse powers of $x$ 
on the left hand side of the above differential equation. 
Upon integrating the above equation under these conditions, we obtain $\dk$ 
to be
\begin{eqnarray}
\dk(x)&=& -\f{3\,i\,b\,f^2}{2\,\Mpl^2}\,
\biggl[{\rm e}^{i\,(\phk/f+\alpha_1)}\,
{\rm e}^{-\pi\,\Mpl^2/(2\,f\,\phi_\ast)}\,
\Gamma\l(1+\f{i\,\Mpl^2}{f\,\phi_\ast},-2\,i\,x\r)\nn\\
& &\,-{\rm e}^{-i\,(\phk/f+\alpha_1)}\,
{\rm e}^{\pi\,\Mpl^2/(2\,f\,\phi_\ast)}\,
\Gamma\l(1-\f{i\,\Mpl^2}{f\,\phi_\ast},-2\,i\,x\r)\biggr],\label{eq:dk-full}
\end{eqnarray}
where $\Gamma(a,x)$ represents the incomplete Gamma function.
In the domain $f\phi_\ast/\Mp^2 \ll 1$, the first term in the above expression 
is exponentially suppressed and hence $\dk$ simplifies to be
\begin{equation}
\dk(x) = \f{3\,i\,b\,f^2}{2\,\Mpl^2}\,
{\rm e}^{-i\,(\phk/f+\alpha_1)}\,
{\rm e}^{\pi\,\Mp^2/2\,f\phi_\ast}\
\Gamma\l(1-\f{i\,\Mpl^2}{f\,\phi_\ast},-2\,i\,x\r).\label{eq:dk}
\end{equation}
This expression for $\dk$ allows us to evaluate the tensor power spectrum and,
we find that in the limit $x\to 0$, it can be expressed as
\begin{equation}
\pt(k)=\pt^0\, \l[1-\dk(0)-\dk^\ast(0)\r]
= \pt^0\,\l[1-\f{f}{\phi_\ast}\,
\delta\ns\,\cos\l(\f{\phk}{f}+\beta_1\r)\r],\label{eq:amm-pt}
\end{equation}
where $\pt^0$ represents the amplitude of the tensor power spectrum which arises 
in the slow roll scenario when the oscillations are absent in the potential and 
is given by
\begin{equation}
\pt^0 = \frac{2\,H_0^2}{\pi^2\,\Mp^2}.\label{eq:amm-pt0}
\end{equation}
The amplitude of the oscillations in the tensor power spectrum prove to be about 
$f/\phi_\ast$ (which is nearly $10^{-3}$, for the best fit values) times smaller 
than the magnitude of the oscillations in the case of scalars.
As in the case of the scalar spectral index, it is convenient to split the 
contribution to the tensor spectral index into a slow roll part and a part which 
is first order in $b$ as $\nt = \nt^0+\nt^{\rm c}$. 
The contribution at the zeroth order in $b$ to the tensor spectral index is 
given by $\nt^0 = -2\,\epsilon_{1}^{\ast}$, which is the standard slow roll result. 
The first order correction to the tensor spectral index can be arrived at using 
the tensor power spectrum~(\ref{eq:amm-pt}) and is found to be~\cite{Meerburg:2014bpa}
\begin{equation}
\nt^{\rm c} 
=-3\,b\,\sqrt{\f{2\,\pi\,f}{\Mp}}\,(2\,\epsilon_1^*)^{3/4}\,
\sin \l(\frac{\phk}{f}+\beta_1\r),\label{eq:amm-ntc}
\end{equation}
which reflects the continued oscillations in the tensor power spectrum.

%%%%%%%%%%%%%%%%%%%%%%%%%%%%%%%%%%%%%%%%%%%%%%%%%%%%%%%%%%%%%%%%%%%%%%%%%%%%%%%

\section{The scalar-scalar-tensor cross-correlation in the Maldacena
formalism}\label{sec:amm-itpfs}

The scalar-scalar-tensor cross-correlation in Fourier space, \viz\/
$\cB_{\cR\cR\gamma}^{m_3n_3}(\vka,\vkb,\vkc)$, evaluated towards the end of 
inflation at the conformal time, say, $\ee$, is defined as
\begin{equation}
\langle\, {\hat \cR}_{\vka}(\eta _{\rm e})\, 
{\hat \cR}_{\vkb}(\eta _{\rm e})\,
{\hat \gamma}_{m_3n_3}^{\vkc}(\eta _{\rm e})\, \rangle 
\equiv \l(2\,\pi\r)^3\, \cB_{\sr\sr\g}^{m_3n_3}(\vka,\vkb,\vkc)\;
\delta^{(3)}\l(\vka+\vkb+\vkc\r).\label{eq:sst-cc}
\end{equation}
For convenience, hereafter, we shall write this correlator as
\begin{equation}
\cB_{\sf \cR\cR\g}(\vka,\vkb,\vkc)
= \l(2\,\pi\r)^{-9/2}\, G_{\sf \cR\cR\g}(\vka,\vkb,\vkc).\label{eq:GABC}
\end{equation}

\par

The scalar-scalar-tensor cross-correlation generated in a given inflationary 
model can be evaluated using the Maldacena formalism~\cite{Maldacena:2002vr,
Sreenath:2013xra}.
The first step in the formalism is to arrive at the third order action describing
the perturbations.
With the action at hand, one can use the standard rules of perturbative quantum 
field theory to arrive at the corresponding three-point function.
The scalar-scalar-tensor cross-correlation 
$G_{\sr\sr\gamma}^{m_3n_3}(\vka,\vkb,\vkc)$, when evaluated 
in the perturbative vacuum, can be written as (see, 
for example, Ref.~\cite{Sreenath:2013xra})
\bea \label{eq:Grrg}
G_{\sr\sr\gamma}^{m_3n_3}(\vka,\vkb,\vkc)
& = & \sum_{C=1}^{3}\; G_{\sr\sr\gamma\,(C)}^{m_3n_3}(\vka,\vkb,\vkc)\nn\\
& = & \Mp^2\; \Pi_{m_3n_3,ij}^{\vkc}\, {\hat n}_{1i}\, {\hat n}_{2j}\; 
\sum_{C=1}^{3}\, \bigl[f_{\ska}(\ee)\, f_{\skb}(\ee)\, h_{\skc}(\ee)\nn\\
& & \times\;\cG_{\sr\sr\gamma}^{C}(\vka,\vkb,\vkc)
+\, {\rm complex~conjugate}\bigr],
\end{eqnarray}
where the quantities $\cG_{\sr\sr\gamma}^{C}(\vka,\vkb,\vkc)$ are described 
by the integrals
\begin{subequations}\label{eqs:cGrrg}
\begin{eqnarray}
\cG_{\sr\sr\gamma}^{1}(\vka,\vkb,\vkc)
&=&-2\, i\; \kaa\, \kbb\, \int_{\ei}^{\ee} 
\d\eta\; a^2\, \epsilon_1\, f_{\ska}^{\ast}\,
f_{\skb}^{\ast}\,h_{\skc}^{\ast},\label{eq:cGrrg1}\\
\cG_{\sr\sr\gamma}^{2}(\vka,\vkb,\vkc)
&=&\f{i}{2}\; \f{\kcc^2}{\kaa\,\kbb}\,
\int_{\ei}^{\ee} \d\eta\; a^2\, \epsilon_1^2\, f_{\ska}^{\prime\ast}\,
f_{\skb}^{\prime\ast}\,h_{\skc}^{\ast},\label{eq:cGrrg2}\\
\cG_{\sr\sr\gamma}^{3}(\vka,\vkb,\vkc)
&=&\f{i}{2}\; \f{1}{\kaa\,\kbb}\,
\int_{\ei}^{\ee} \d\eta\; a^2\, \epsilon_1^2\, 
\l[\kaa^2\,f_{\ska}^{\ast}\,f_{\skb}^{\prime\ast}
+\kbb^2\,f_{\ska}^{\prime\ast}\,f_{\skb}^{\ast}\r]\,
h_{\skc}^{\prime\ast}.\label{eq:cGrrg3} 
\end{eqnarray}
\end{subequations}
The lower limit of the integrals, \viz $\ei$, denotes a sufficiently 
early time at which the initial conditions are imposed on the modes 
when they are well inside the Hubble radius.
The upper limit $\ee$ denotes a suitably late time which can, for 
instance, be conveniently chosen to be a time close to the end of 
inflation.
Note that for a given wavevector ${\bm k}$, ${\hat {\bm n}}$ denotes 
the unit vector ${\hat {\bm n}}={\bm k}/k$.
Hence, the quantities ${\hat n}_{1i}$ and ${\hat n}_{2i}$ represent 
the components of the unit vectors ${\hat {\bm n}}_{1}={\bm k}_1/\kaa$ 
and ${\hat {\bm n}}_{2}={\bm k}_2/\kbb$ along the $i$-spatial direction.

\par

As in the case of the scalar bi-spectrum, it proves to be convenient to
introduce a dimensionless non-Gaussianity parameter, say, $\cnls(\vka,
\vkb,\vkc)$, to reflect the amplitude of the scalar-scalar-tensor 
three-point function.
It can be defined as a suitable ratio of the three-point function and the 
scalar and tensor power spectra as follows~\cite{Sreenath:2013xra,
Sreenath:2014nka}:
\begin{eqnarray}
\cnls(\vka,\vkb,\vkc) 
&=& -\f{4}{\l(2\,\pi^2\r)^2}\,
\l[k_1^3\, k_2^3\, k_3^3\; G_{\cR\cR\g}^{m_3n_3}(\vka,\vkb,\vkc)\r]\nn\\
& &\times\; {\l(\Pi_{m_3n_3,{\bar m}{\bar n}}^{\vkc}\r)}^{-1}\;\,
\biggl\{\l[k_1^3\; {\mathcal P}_{_{\rm S}}(k_2)
+k_2^3\; {\mathcal P}_{_{\rm S}}(k_1)\r]\; 
{\mathcal P}_{_{\rm T}}(k_3)\biggr\}^{-1}.\label{eq:cnls}
\end{eqnarray}

%%%%%%%%%%%%%%%%%%%%%%%%%%%%%%%%%%%%%%%%%%%%%%%%%%%%%%%%%%%%%%%%%%%%%%%%%%%%%%%

\section{Analytical template for the scalar-scalar-tensor cross-correlation}
\label{sec:amm-sstcc}

In this section, we shall first arrive at an analytical expression for the
scalar-scalar-tensor three-point function.
Then, in order to illustrate the accuracy of the analytical results, we shall
compare them with the exact numerical results.

%%%%%%%%%%%%%%%%%%%%%%%%%%%%%%%%%%%%%%%%%%%%%%%%%%%%%%%%%%%%%%%%%%%%%%%%%%%%%%%

\subsection{Analytical evaluation of the three-point function}

Among the three different contributions to the scalar-scalar-tensor three-point 
function, the term $\cG_{\sr\sr\gamma}^{1}(\vka,\vkb,\vkc)$ 
[cf.~Eq.~(\ref{eq:cGrrg1})] is linear in the first slow roll parameter $\epsilon_1$ 
and hence it dominates over the other two terms [cf.~Eqs.~(\ref{eq:cGrrg2}) 
and~(\ref{eq:cGrrg3})], both of which are quadratic in $\epsilon_1$. 
The term $\cG_{\sr\sr\gamma}^{1}(\vka,\vkb,\vkc)$ can be decomposed into a slow 
roll part, which is zeroth order in $b$, and terms involving $b$. 
The contribution when $b=0$ corresponds to the standard slow roll 
result and it can be easily evaluated using the de Sitter modes to 
be~\cite{Sreenath:2014nka}
\begin{eqnarray}\label{eq:b0}
G_{\sr\sr\gamma}^{1(0)}(\vka,\vkb,\vkc) 
&=& \Pi_{m_3n_3,ij}^{\vkc}\, {\hat n}_{1i}\, {\hat n}_{2j}\; 
\frac{H_0^4\, k_1\, k_2}{4\,\Mpl^4\,\epsilon_1^{\ast}\,
\l(k_1\,k_2\,k_3\r)^3}
\biggl[-\kt+\frac{k_1\,k_2+k_2\,k_3+k_3\,k_1}{\kt}\nn\\
& &+\,\frac{k_1\,k_2\,k_3}{\kt^2}\biggr],
\end{eqnarray}
where $\kt=k_1+k_2+k_3$, and we have suppressed the indices $m_3$ and $n_3$
on $G_{\sr\sr\gamma}(\vka,\vkb,\vkc)$ for convenience.

\par

Let us now turn to the contributions involving $b$.
As we have discussed, we shall ignore terms which are of higher order in $b$ and 
focus only on the contributions that are linear in~$b$.
Even amongst the various terms which are linear in $b$, we shall further restrict
ourselves to terms which are of the leading order in $f\,\phi_\ast/\Mp^2$.
In the case of the scalar bi-spectrum, the dominant contribution arises due to
a term dependent on ${\dot \epsilon}_2$, which grows to be quite large in the 
axion monodromy model. 
This in turn boosts the scalar bi-spectrum and the corresponding non-Gaussianity 
parameter to rather significant values~\cite{Flauger:2010ja}.
Moreover, since ${\dot \epsilon}_2$ becomes large, it proves to be sufficient to 
work with the de Sitter modes to evaluate the dominant contribution.   
In contrast, in the case of the scalar-scalar-tensor cross-correlation, apart from
the correction to the slow roll parameter $\epsilon_1^{\ast}$ (\viz\/ $\epsilon_1^c$), 
we have to take into account the modification to the de Sitter modes, which are 
quantified by $\ck$ and $\dk$.
It should be clear that, at the linear order in $b$, there arise four contributions 
due to $c_k$ and $\dk$, two from the $c_k(x)$ and $\dk(x)$ inside the integral 
[cf. Eq.~(\ref{eq:cGrrg1})] and two others due to the terms $\ck(0)$ and $\dk(0)$
outside [cf. Eq.~(\ref{eq:Grrg})].
One finds that $\dk/\ck\sim f/\phi_{\ast}$, which is a small quantity.
Therefore, one can actually ignore the terms involving $\dk$ and retain only those
containing~$\ck$.
Under these conditions, at the first order in $b$, we can write the expression for 
the scalar-scalar-tensor three-point function as
\begin{eqnarray}\label{eq:sst}
G_{\sr\sr\gamma}^{1(1)}(\vka,\vkb,\vkc) 
&=& \Mpl^2\,\Pi_{m_3n_3,ij}^{\vkc}\, {\hat n}_{1i}\, {\hat n}_{2j}\;
\Biggl\{-2\,i\,\kaa\,\kbb\,\biggl[\fka^+(0)\,\fkb^+(0)\,\gkc^+(0)\,
\int_{-\infty}^{0}{\rm d}\eta\;a^2\nn\\
& &\times\,\l(\epsilon_1^*\,\cka^*\,\fka^+\,\fkb^{+*}\,\gkc^{+*}
+\epsilon_1^*\,\ckb^*\,\fka^{+*}\,\fkb^{+}\,\gkc^{+*}
+\epsilon_1^c\,\fka^{+*}\,\fkb^{+*}\,\gkc^{+*}\r)\nn\\
& &+\,\l(\cka(0)\,\fka^{+*}(0)\,\fkb^{+}(0)\,\gkc^{+}(0)
+ \ckb(0)\,\fka^{+}(0)\,\fkb^{+*}(0)\,\gkc^{+}(0)\r)\nn\\
& &\times\;
\int_{-\infty}^{0}{\rm d}\eta\; a^2\,\epsilon_1^*\,
\fka^{+*}\,\fkb^{+*}\,\gkc^{+*}\biggr]+{\rm complex~conjugate}\Biggr\}.
\end{eqnarray}

\par

Let us first consider the term containing $\epsilon_1^c$ in the above expression.
At the linear order in $b$ and $f\,\phi_\ast/\Mpl^2$, we have
\begin{eqnarray}\label{eq:G-eps}
G_{\sr\sr\gamma}^{1(1a)}(\vka,\vkb,\vkc) 
&=& \Pi_{m_3n_3,ij}^{\vkc}\, {\hat n}_{1i}\, {\hat n}_{2j}\;
\Biggl[\frac{H_0^4}{8\,\Mp^4\, \epsilon_1^{*2}}\,
\f{-i\,\kaa\,\kbb}{(\kaa\,\kbb\,\kcc)^3}\,
\int_{-\infty}^0\,\frac{\d\eta}{\eta^2}\,\epsilon_1^c\,
\Biggl(1-i\kt\eta \nn\\
& & -\l(\kaa\,\kbb+\kbb\,\kcc+\kcc\,\kaa\r)\,\eta^2
+i\,\kaa\,\kbb\,\kcc\eta^3\Biggr)\,{\rm e}^{i\,\kt\,\eta}\nn\\
& & +\,{\rm complex~conjugate}\Biggr],
\end{eqnarray}
where we have used the expressions~(\ref{eq:amm-s-f}) and~(\ref{eq:amm-s-g}) for 
the modes $\fk$ and $\gk$ in~Eqs.~(\ref{eq:Grrg}) and~(\ref{eq:cGrrg1}).
We can use the expressions (\ref{eq:eps1c}) and (\ref{eq:amm-phi0}) and substitute 
$x=-\kt\eta$ for performing the above integrals. 
Each of these integrals are found to be of the following form:
\begin{equation}
I_1(k_1,k_2,k_3,X_{\rm res},f) 
= \int_{0}^{\infty}{\rm d}x\,q(x)\,{\rm e}^{-i\,x}\,
\l\{{\rm e}^{i\,\l[(\phi_{\kt}/f)+X_{\rm res}\,{\rm ln}\,x\r]}
+{\rm e}^{-i\,\l[(\phi_{\kt}/f)+X_{\rm res}\,{\rm ln}\,x\r]}\r\},
\end{equation}
where $q\l(x\r)$ is some polynomial function of $x$. 
The two terms in the above integral can be expressed in terms of the Gamma functions. 
However, we find that, for small $f\,\phi_\ast/\Mp^2$, the contribution due to the 
second term is exponentially suppressed when compared to the first term and hence
can be ignored. 
Under this assumption, we can evaluate the integrals in Eq.~(\ref{eq:G-eps})  to 
obtain
\begin{eqnarray}
G_{\sr\sr\gamma}^{1(1a)}(\vka,\vkb,\vkc) 
&=& \Pi_{m_3n_3,ij}^{\vkc}\, 
{\hat n}_{1i}\, {\hat n}_{2j}\; 
\f{H_0^4}{4\,\Mp^4\, \epsilon_1^*}\, 
\f{3\,b\,\sqrt{2\,\pi}}{\sqrt{X_{\rm res}}}\, 
\f{\kaa\,\kbb}{(\kaa\,\kbb\,\kcc)^3}\nn\\
& & \times\, \Biggl\{\f{\kt}{1+X_{\rm res}^2}\,
\Biggl[-\sin\l(\f{\phi_{\kt}}{f}+\beta_2\r) 
+\f{1}{X_{\rm res}}\,\cos\l(\f{\phi_{\kt}}{f}+\beta_2\r)\Biggr] \nn\\
& & -\,\f{\kt}{X_{\rm res}}\,\cos\l(\f{\phi_{\kt}}{f}+\beta_2\r) 
+ \f{\kaa\,\kbb + \kbb\,\kcc + \kcc\,\kaa}{\kt}\,
\sin\l(\f{\phi_{\kt}}{f}+\beta_2\r) \nn\\
& & +\, \f{\kaa\,\kbb\,\kcc}{\kt^2}\,
\Biggl[X_{\rm res}\,\cos\l(\f{\phi_{\kt}}{f}+\beta_2\r) 
+ \sin\l(\f{\phi_{\kt}}{f}+\beta_2\r)\Biggr]\Biggr\},
\label{eq:g1-eps}
\end{eqnarray}
where the phase factor $\beta_2$ is given by 
$\beta_2 = X_{\rm res}\,{\rm ln}\,X_{\rm res}-X_{\rm res}-\pi/4$.

\par 

Let us now consider the terms containing $\ck$ in Eq.~(\ref{eq:sst}). 
We shall consider terms involving both $\ck(0)$ as well as $\ck(\eta)$.
We can substitute the expressions for the modes $f_k^{+}$ and $g_k^{+}$ 
[cf.~Eqs.~(\ref{eq:de-fk}) and~(\ref{eq:de-hk})] to obtain the contributions
due to these terms to be
\begin{eqnarray}\label{eq:G-ck}
G_{\sr\sr\gamma}^{1(1b)}(\vka,\vkb,\vkc) 
&=& \Pi_{m_3n_3,ij}^{\vkc}\, {\hat n}_{1i}\, {\hat n}_{2j}\;
\Biggl\{\f{i\,H_0^4}{8\,\Mp^4\,\epsilon_1^*}\,
\frac{\kaa\,\kbb}{(\kaa\,\kbb\,\kcc)^3}\,\biggl[\int_{-\infty}^0\,
\frac{\d\eta}{\eta^2}\,\cka^*(\eta)\nn\\
& &\times\,\biggl(\underbrace{1-i\,\ka\,\eta}_\textrm{\RNum{1}} 
+\underbrace{(\kaa\,\kbb-\kbb\,\kcc+\kcc\,\kaa)\,\eta^2
-i\,\kaa\,\kbb\,\kcc\,\eta^3}_\textrm{\RNum{2}}\biggr)\, 
{\rm e}^{i\,\ka\,\eta} \nn\\
& &+\, \cka(0)\,\int_{-\infty}^0\,\frac{\d\eta}{\eta^2}\,
\biggl(\underbrace{1-i\,\kt\,\eta}_\textrm{\RNum{1}}
-\underbrace{(\kaa\,\kbb+\kbb\,\kcc+\kcc\,\kaa)\,\eta^2}_\textrm{\RNum{2}}\nn\\ 
& &+\,\underbrace{i\,\,\kaa\,\kbb\,\kcc\,\eta^3}_\textrm{\RNum{2}}\biggr)\, 
{\rm e}^{i\,\kt\,\eta}\biggr]+{\rm complex~conjugate}\Biggr\} \nn\\
& & +\,{\rm a~similar~term~with}~\kaa~{\rm and}~\kbb~{\rm exchanged},
\end{eqnarray}
where $\ka = \kt - 2\,\kaa$.
Let us first consider the integrals which have been highlighted 
as~$(\textrm{\RNum{1}})$ in the above equation.
The integrals involving $\cka^\ast(\eta)$ are found to diverge as $\eta \to 0$
[note that $\ck(\eta)$ is given by Eq.~(\ref{eq:ck})].
However, as we shall soon see, their complete contribution to the three-point 
function proves to be finite in the limit.
Therefore, we initially set the upper limit of the integrals to be, say, 
$\ee$ (which denotes the conformal time at the end of inflation), and 
eventually consider the $\ee\to 0$ limit.
We also evaluate the integrals containing $\cka(0)$ in the same fashion.
Thereafter, we combine all the integrals marked as $(\textrm{\RNum{1}})$,
add the resultant expressions to their complex conjugates, and take the
$\ee\to 0$ limit to finally arrive at the corresponding contribution to 
the scalar-scalar-tensor cross-correlation.
We find that the contributions due to the terms marked as $(\textrm{\RNum{1}})$
can be written as
\begin{eqnarray}
G_{\sr\sr\gamma}^{1(1b\textrm{\RNum{1}})}(\vka,\vkb,\vkc) 
&=&\Pi_{m_3n_3,ij}^{\vkc}\, {\hat n}_{1i}\, {\hat n}_{2j}\;
\Biggl\{\f{i\,H_0^4}{8\,\Mp^4\,\epsilon_1^*}\,
\frac{\kaa\,\kbb}{(\kaa\,\kbb\,\kcc)^3}\nn\\
& &\times\,\Biggl[\int_{-\infty}^0\,\frac{\d\eta}{\eta^2}\,\cka^*(\eta)\,
\l(1-i\,\ka\,\eta\r)\,{\rm e}^{i\,\ka\,\eta}\nn\\
& &+\,\cka(0)\int_{-\infty}^0\,\frac{\d\eta}{\eta^2}\,
\l(1-i\,\kt\,\eta\r)\, {\rm e}^{i\,\kt\,\eta}\Biggr]
+{\rm complex~conjugate}\Biggr\}\nn\\
&=&\f{H_0^4}{4\,\Mp^4\, \epsilon_1^\ast}\, 
\f{3\,b\,\sqrt{2\,\pi}}{\sqrt{X_{\rm res}}}\, 
\f{\kaa\,\kbb}{(\kaa\,\kbb\,\kcc)^3}\,
\l(\kbb+\kcc\r)\,\sin\l({\frac{\phka}{f}+\beta_1}\r),
\label{eq:g1-ckI}
\end{eqnarray}
where, to obtain the final result, we have made use of the 
expression~(\ref{eq:ck0}) for $\cka(0)$.

\par

We can now consider the integrals which have been indicated as 
$(\textrm{\RNum{2}})$ in Eq~(\ref{eq:G-ck}).
We switch to the variable $x = -\kaa\eta$ and substitute for $\cka(\eta)$ 
from Eq.~(\ref{eq:ck}). 
Note that the expression for $\cka(\eta)$ involves the incomplete Gamma
function [cf.~Eq.~(\ref{eq:ck})]. 
These terms contain integrals of the following form:
\begin{equation}
I_2\l(k_1,k_2,k_3,X_{\rm res}\r) 
= \int_0^{\infty}\d x\, u(x)\, {\rm e}^{v(x)}\,
\Gamma\l(1+i\,X_{\rm res},2\,i\,x\r),
\end{equation}
where $u(x)$ and $v(x)$ are some polynomial functions of $x$ and 
$\Gamma\l(1+i\,X_{\rm res},2\,i\,x\r)$ is the incomplete Gamma 
function~\cite{Gradshteyn:2007}.
We find that these integrals can be evaluated if we make use of the integral
representations for the incomplete Gamma function and interchange the order
of the integrals as follows:
\begin{equation}
\int_0^{\infty}\d x\,u(x)\,{\rm e}^{v(x)}\,
\int_{2\,i\,x}^{\infty}\d y\,y^{i\,X_{\rm res}}\, {\rm e}^{-y}
= \int_0^{\infty}\d y\,y^{i\,X_{\rm res}}\,
{\rm e}^{-y}\,\int_0^y\frac{\d p}{2\,i}\,u\left(\frac{p}{2\,i}\right)\,
{\rm e}^{v(p/2\,i)},
\label{eq:integ-exchng}
\end{equation}
where we have set $p = 2\,i\,x$. 
The complete contribution due to the terms marked as $(\textrm{\RNum{2}})$ is
found to be
\begin{eqnarray}
G_{\sr\sr\gamma}^{1(1b\textrm{\RNum{2}})}(\vka,\vkb,\vkc) 
&=& -\Pi_{m_3n_3,ij}^{\vkc}\, {\hat n}_{1i}\, {\hat n}_{2j}\,
\f{H_0^4}{4\,\Mp^4\, \epsilon_1^\ast}\, 
\f{3\,b\,\sqrt{2\,\pi}}{\sqrt{X_{\rm res}}}\, 
\f{\kaa\,\kbb}{(\kaa\,\kbb\,\kcc)^3} \nn\\
& & \times\, \Biggl(\f{\kaa\,\kbb+\kbb\,\kcc+\kcc\,\kaa}{2\,\kt}\sin\l(\f{\phka}{f}+\beta_1\r)
+\f{\kaa\,\kbb\,\kcc}{2\,\kt^2}\, \sin\l(\f{\phka}{f}+\beta_1\r)\nn\\
& & +\,\f{\kaa\,\kbb-\kbb\,\kcc+\kcc\,\kaa}{2\ka}\,
\Biggl\{\f{1}{1+\l(\ka/2\,\kaa\r)}\,\nn\\
& &\times\,\sin\l[\f{\phka}{f}+\beta_1  - X_{\rm res}\,\ln\l(1+\f{\ka}{2\,\kaa}\r)\r]
-\sin\l(\f{\phka}{f}+\beta_1\r)\Biggr\}\nn\\
& & +\, \f{\kbb\,\kcc}{4\,\ka}\,\l(1+\f{\ka}{2\kaa}\r)^{-2}\, 
\Biggl\{\sin\l[\f{\phka}{f}+\beta_1 - X_{\rm res}\,\ln\l(1+\f{\ka}{2\kaa}\r)\r]\nn\\
& & +\, X_{\rm res}\, \cos\l[\f{\phka}{f}+\beta_1 
- X_{\rm res}\,\ln\l(1+\f{\ka}{2\,\kaa}\r)\r]\Biggr\}\nn\\
& & +\,\f{\kaa\,\kbb\,\kcc}{2\,\ka^2}\,
\Biggl\{\f{1}{1+\l(\ka/2\,\kaa\r)}\,\sin\l[\f{\phka}{f}+\beta_1 - X_{\rm res}\,
\ln\l(1+\f{\ka}{2\,\kaa}\r)\r] \nn\\
& & -\,\sin\l(\f{\phka}{f}+\beta_1\r)\Biggr\} \Biggr)\nn \\
& & +\,{\rm a~similar~term~with}\; \kaa\; {\rm and}\; \kbb \; {\rm exchanged}.
\label{eq:g1-ckII}
\end{eqnarray}

\par

The sum of the contributions~(\ref{eq:g1-eps}), (\ref{eq:g1-ckI}) 
and~(\ref{eq:g1-ckII}) together with the contribution~(\ref{eq:b0}) 
gives the complete contribution to the scalar-scalar-tensor 
cross-correlation under the approximations we have worked with.

%%%%%%%%%%%%%%%%%%%%%%%%%%%%%%%%%%%%%%%%%%%%%%%%%%%%%%%%%%%%%%%%%%%%%%%%%%%%%%%

\subsection{Comparison with the numerical results}\label{subsec:amm-comp}

\begin{figure}[!h]
\begin{center}
\includegraphics[width=7.5cm]{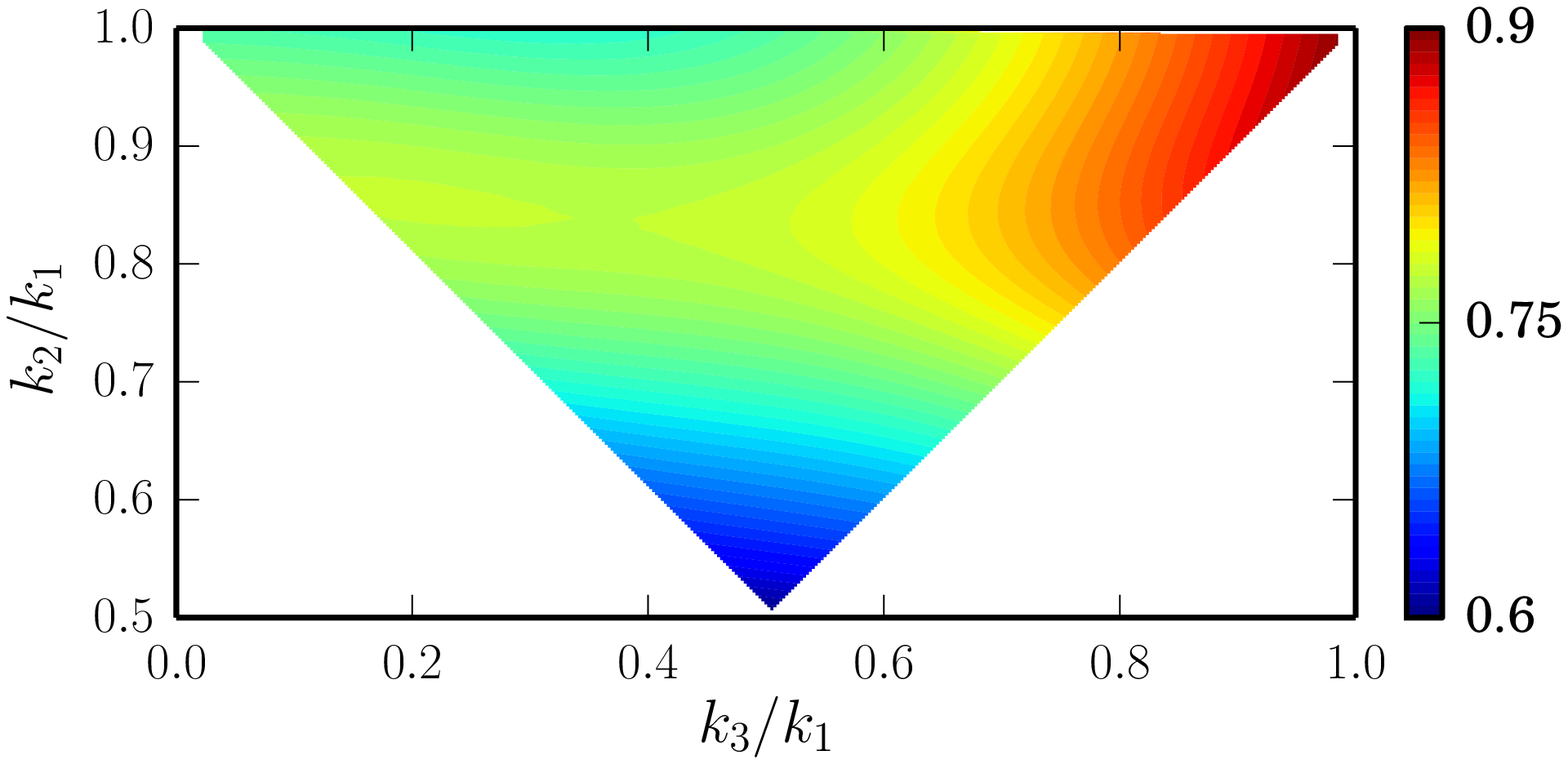}
\includegraphics[width=7.5cm]{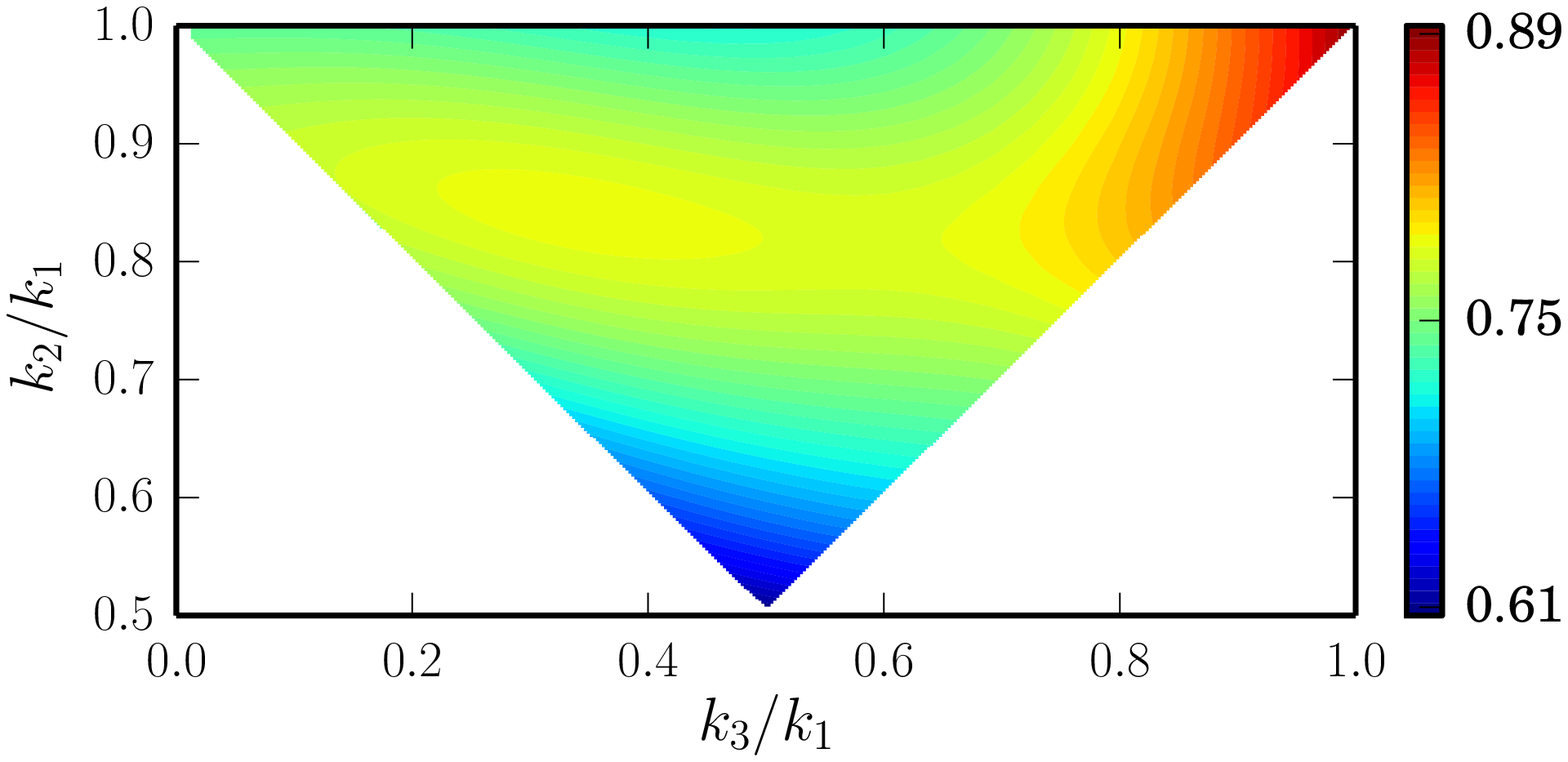}\\
\includegraphics[width=7.5cm]{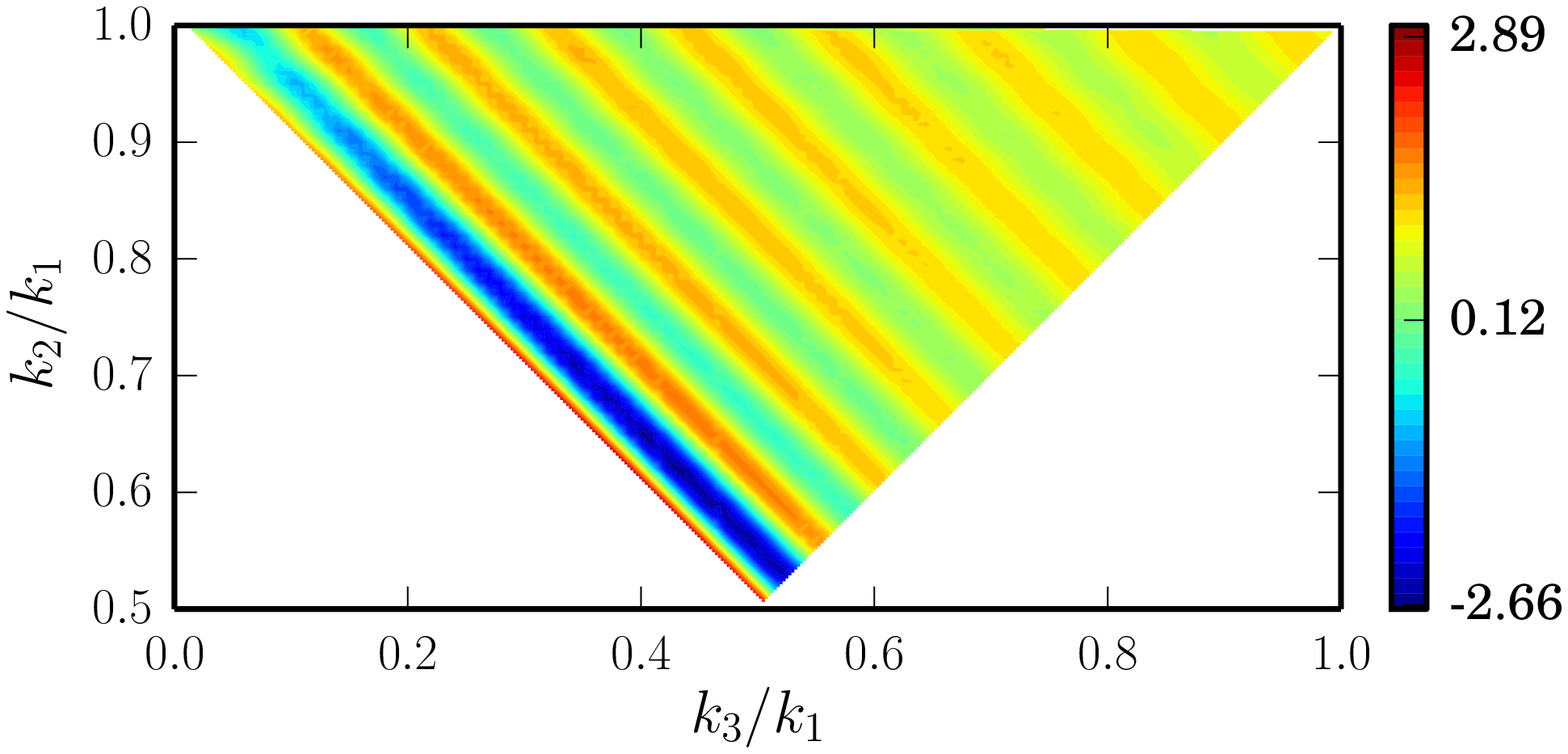}
\includegraphics[width=7.5cm]{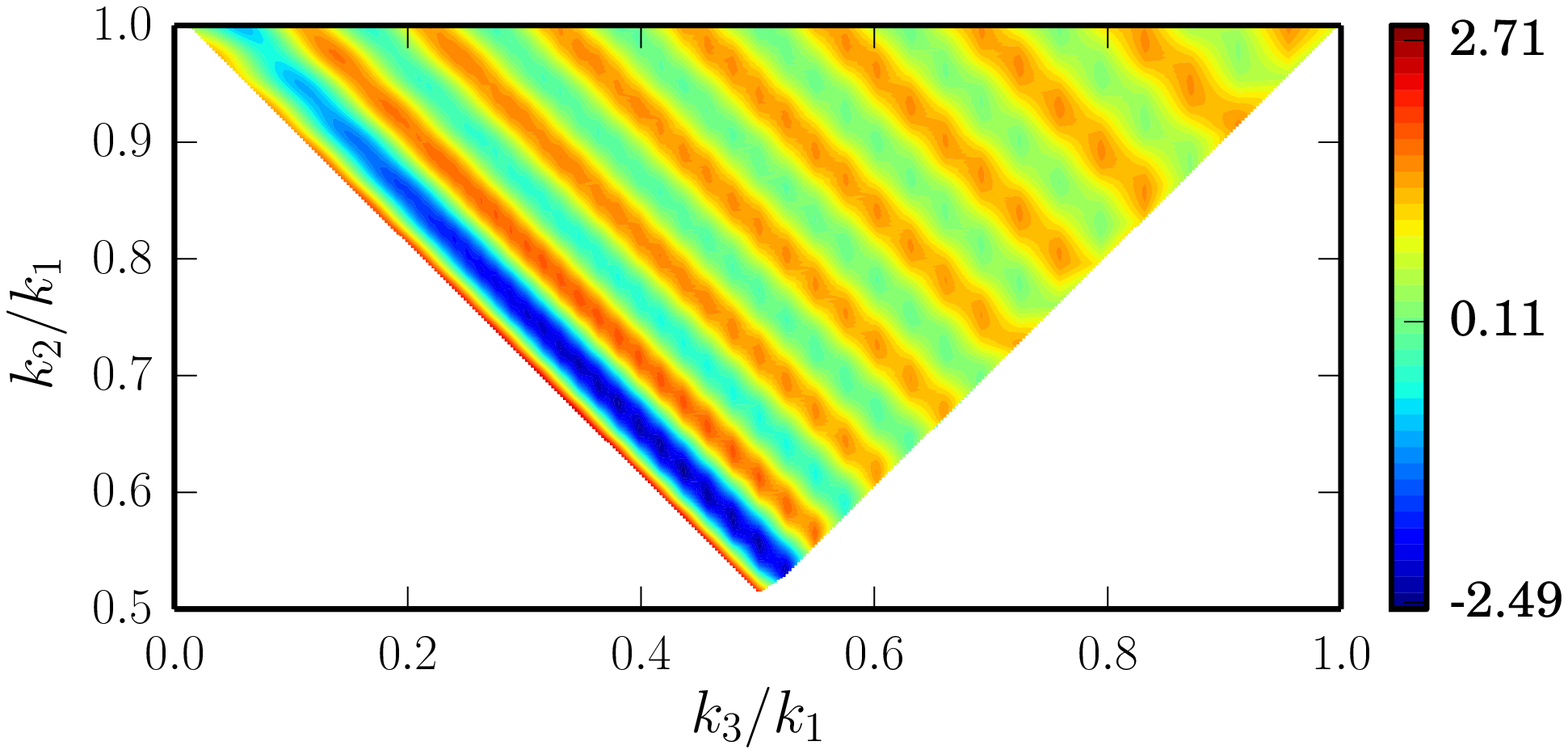}
\end{center}
\caption{A comparison of the analytical results (on the left) with the numerical
results (on the right) for the non-Gaussianity parameter $\cnls(\vka,\vkb,\vkc)$ 
characterizing the scalar-scalar-tensor three-point function.
We have plotted the results for two sets of values of the parameters involved.
We have chosen $\mu/\Mp = 2.512\times 10^{-10}$ in arriving at all these figures.
The results in the top row correspond to $b = 1.063\times 10^{-2}$ and $f/\Mpl 
= 7.6346\times 10^{-3}$, which are values that lead to the best fit to the
Planck data~\cite{Meerburg:2013dla}.
The results in the bottom row correspond to $b = 5.0\times 10^{-2}$ and $f/\Mpl 
= 7.6346\times 10^{-4}$.
It is clear that the analytical results match the numerical results very well 
for both these sets of values.
In fact, we find that the match is always better than $7\%$ over the range of 
wavenumbers that we have considered.
Note that, as expected, the non-Gaussianity parameter $\cnls(\vka,\vkb,\vkc)$ 
exhibits more oscillations for smaller values of $f$, as is illustrated by the
figures in the bottom row.
Also, the strength of these oscillations is more for larger values of~$b$.}
\label{fig:comparison-numerical}
\end{figure}

In order to illustrate the accuracy of our analytical calculations, we shall 
now compare our analytical results that have been arrived at under certain 
approximations with the exact results obtained numerically.
We have obtained the numerical results using a code we had developed 
earlier (for details about the code, see Ref.~\cite{Sreenath:2013xra}).
In fact, we shall compare the results for the corresponding non-Gaussianity 
parameter $\cnls$ [cf.~Eq.~(\ref{eq:cnls})].
In Fig.~\ref{fig:comparison-numerical}, we have plotted the analytical and
the exact numerical results for two sets of values of the parameters involved.
We have chosen values for the parameters such that the approximations we have
worked with are valid.
%%%%%%%%%%%%%%%%%%%%%%%%%%%%%%%%%%%%%%%%%%%%%%%%%%%%%%%%%%%%%%%%%%%%%%%%%%%%%%%
%%%%%%%%%%%%%%%%%%%%%%%%%%%%%%%%%%%%%%%%%%%%%%%%%%%%%%%%%%%%%%%%%%%%%%%%%%%%%%%
It is evident from the figure that the analytical results match the numerical
ones quite well. 

%%%%%%%%%%%%%%%%%%%%%%%%%%%%%%%%%%%%%%%%%%%%%%%%%%%%%%%%%%%%%%%%%%%%%%%%%%%%%%%

\section{The squeezed limit and the consistency relation}\label{sec:sqlcr}

In this section, we shall discuss the behavior of the scalar-scalar-tensor 
three-point function in the so-called squeezed limit.
This limit corresponds to the situation where one of the three wavenumbers 
involved is much smaller than the other two.
In such a limit, it is well known that the three-point functions can be written 
in terms of the two-point functions through a relation known as the consistency 
condition~\cite{Maldacena:2002vr,Creminelli:2004yq,Cheung:2007sv,
RenauxPetel:2010ty,Ganc:2010ff,Creminelli:2011rh,Chialva:2011iz,
Schalm:2012pi,Pajer:2013ana,Sreenath:2014nka}. 
This condition primarily arises due to the fact that the amplitude of the 
long wavelength scalar and tensor modes freeze on supper Hubble scales and 
hence can be treated as a background as far as the smaller wavelength modes 
are concerned. 
In the case of the scalar-scalar-tensor three-point function, when the 
wavenumber of the tensor mode is considered to be much smaller than the 
two scalar modes, it is found that the three-point function can be 
completely expressed in terms of the scalar and tensor power spectra as 
follows (see, for instance, Refs.~\cite{Maldacena:2002vr,Sreenath:2014nka}):
\begin{eqnarray} 
\langle\, {\hat \cR}_{\vka}(\eta _{\rm e})\, 
{\hat \cR}_{\vkb}(\eta _{\rm e})\,
{\hat \gamma}_{m_3n_3}^{\vkc}(\eta _{\rm e})\,\rangle_{\kcc}
&=& -\,\frac{(2\,\pi)^{5/2}}{4\,\kaa^3\,\kcc^3}\,
\l(\frac{\ns-4}{8}\r)\,\ps(\kaa)\,\pt(\kcc)\nn\\
& &\times\, \Pi_{m_3n_3,ij}^{\vkc}\, {\hat n}_{1i}\, 
{\hat n}_{1j}\, \delta^3(\vka+\vkb),
\end{eqnarray}
where we have considered $\vkc$ to be the squeezed mode.
The above condition can essentially be expressed as
\begin{eqnarray} 
k^3\,\kcc^3\,G_{\sr\sr\gamma}^{m_3n_3}(\vk,-\vk,\vkc)
&=& -\Pi_{m_3n_3,ij}^{\vkc}\, {\hat n}_{i}\, {\hat n}_{j}\,
\frac{(2\,\pi)^{4}}{4}\, \l(\frac{\ns-4}{8}\r)\,\ps(k)\,\pt(\kcc),
\label{eq:conrel}
\end{eqnarray}
with the limit $\kcc\to 0$ kept in mind.
In what follows, using the analytical results we have obtained for the power 
spectra and the 
scalar-scalar-tensor cross-correlation, we shall explicitly show that such  
a consistency relation is indeed satisfied in the axion monodromy model.

\par

Let us now consider the squeezed limit of the three-point function we have
arrived at analytically.
In the limit $\vka = -\vkb = \vk$ and $\kcc \rightarrow 0$, at the leading 
order in $X_{\rm res}$, we find that the three-point 
function at the order $b$ [\ie\/ the sum of the contributions~(\ref{eq:g1-eps}),  
(\ref{eq:g1-ckI}) and~(\ref{eq:g1-ckII})] reduces to
\begin{eqnarray}
k^3\,\kcc^3\, G_{\sr\sr\gamma}^{1(1)}(\vk,-\vk,\vkc) 
&=& \Pi_{m_3n_3,ij}^{\vkc}\, {\hat n}_{i}\, {\hat n}_{j}\,
\f{H_0^4}{8\, i\, \Mp^4\, \epsilon_1^\ast}\, 
\f{3\,b\,\sqrt{2\,\pi}}{\sqrt{X_{\rm res}}}\,\f{k}{\ka} \nn\\
& & \times\, \Biggl({\rm e}^{i\,\phi_{k}/f}\,
\l\{\frac{\l[1+\l(\ka/2\,k\r)\r]^{-i\,X_{\rm res}}}{1+\l(\ka/2\,k\r)}-1\r\}\nn\\
& & -\,{\rm e}^{-i\,\phi_{k}/f}\,
\l\{\frac{\l[1+\l(\ka/2\,k\r)\r]^{i\,X_{\rm res}}}{1+\l(\ka/2\,k\r)}-1\r\}\Biggr),
\end{eqnarray}
where, recall that, $\ka=\kt-2\,\kaa$.
Hence, in the squeezed limit, $\ka \rightarrow 0$. 
Therefore, we can expand the terms $(1+\ka/2\,k)^{\pm i\,X_{\rm res}}$ in the above 
equation up to the first order in $\ka$ to obtain 
the following expression for the three-point function:
\begin{eqnarray}
k^3\,\kcc^3\, G_{\sr\sr\gamma}^{1(1)}(\vk,-\vk,\vkc) 
&=& -\Pi_{m_3n_3,ij}^{\vkc}\, {\hat n}_{i}\, {\hat n}_{j}\;
\frac{3\,b\,H_0^4\,\sqrt{2\,\pi}}{8\,\Mp^4\,\epsilon_1^\ast}\,
X_{\rm res}^{1/2}\,\cos\left(\frac{\phk}{f}\right).\label{eq:amm-rrg-sl}
\end{eqnarray}
We should mention that, in arriving at this expression, we have ignored a 
$k$-independent phase.

\par

Let us now turn to the right hand side of the relation~(\ref{eq:conrel}).
Up to the linear order in~$b$, we can have four combinations of the various
terms, given by
\begin{eqnarray}
\l(\frac{\ns-4}{8}\r)\,\ps(k)\,\pt(\kcc) 
&\simeq& \l(\frac{\ns^0-4}{8}\r)\,\ps^0(k)\,\pt^0(\kcc)
+\l(\frac{\ns^0-4}{8}\r)\,\ps^{\rm c}(k)\,\pt^0(\kcc) \nn\\
& &+\,\l(\frac{\ns^{\rm 0}-4}{8}\r)\,\ps^0(k)\,\pt^{\rm c}(\kcc)\,+\,
\l(\frac{\ns^{\rm c}}{8}\r)\,\ps^0(k)\,\pt^0(\kcc),\quad
\end{eqnarray}
where $\ps^0(k)$ and $\pt^0(k)$ and $\ns^0$ are the scalar and the tensor 
power spectra and the scalar spectral index, respectively, which arise in 
the absence of the oscillations in the axion monodromy model.  
Note that $\pt^{\rm c}(k)$ involves terms of order $\dk$ and, as we have 
discussed before, these terms are of lower order when compared to the 
other terms involving $\ck$.
Hence, for consistency, we can ignore the contribution due to 
$\pt^{\rm c}(k)$ in the above equation. 
Therefore, we finally obtain that
\begin{eqnarray}
\l(\frac{\ns-4}{8}\r)\,\ps(k)\,\pt(\kcc) 
&\simeq& \l(\frac{\ns^0-4}{8}\r)\,\ps^0(k)\,\pt^0(\kcc)
+\l(\frac{\ns^0-4}{8}\r)\,\ps^{\rm c}(k)\,\pt^0(\kcc)\nn\\
& & +\, \l(\frac{\ns^{\rm c}}{8}\r)\,\ps^0(k)\,\pt^0(\kcc).
\end{eqnarray}

\par

The first term in the above expression is the slow roll term for which the 
consistency relation involving the $b=0$ contribution to the 
scalar-scalar-tensor cross correlation, \viz\/ 
$G_{\sr\sr\gamma}^{1(0)}(\vka,\vkb,\vkc)$, can be verified 
easily~\cite{Kundu:2013gha,Sreenath:2014nka}.
On evaluating the remaining two terms, we find that the one involving $\ns^{\rm c}$ 
is of leading order, as the other term is suppressed relative to this by a factor 
of $X_{\rm res}$. 
Hence, at the leading order in $b$, we can replace 
$\l[\l(\ns-4\r)/8\r]\,\ps(k)\,\pt(\kcc)$ by 
$\l(\ns^{\rm c}/8\r)\,\ps^0(k)\,\pt^0(\kcc)$ in Eq.~(\ref{eq:conrel}). 
Upon making this replacement, the consistency relation at the linear order in $b$ 
can be written as
\begin{eqnarray}
k^3\,\kcc^3\,G_{\cR\cR\gamma}^{1(1)}(\vk,-\vk,\vkc)
=-\Pi^{\vkc}_{m_3n_3,ij}\, {\hat n}_{1i}\, {\hat n}_{1j}\,
\f{(2\,\pi)^{4}}{4}\, \l(\f{\ns^{\rm c}}{8}\r)\, \ps^0\, \pt^0.
\end{eqnarray}
Now, on substituting the expression for $\ns^c$ [cf. Eqs.~(\ref{eq:amm-nsc})]
and the slow roll amplitudes for the scalar and tensor power spectra 
[cf. Eqs.~(\ref{eq:amm-ps0}) and~(\ref{eq:amm-pt0})] in the above
expression, we find that the resultant expression is the same as that obtained 
in Eq.~(\ref{eq:amm-rrg-sl}), up to a $k$-independent phase. 
This implies that the consistency relation is valid in the axion monodromy model 
even in the presence of persistent oscillations in the two as well as the 
three-point functions~\cite{Sreenath:2014nka}.

%%%%%%%%%%%%%%%%%%%%%%%%%%%%%%%%%%%%%%%%%%%%%%%%%%%%%%%%%%%%%%%%%%%%%%%%%%%%%%%

\section{Discussion}\label{sec:d}

The axion monodromy model is described by a linear potential with small 
periodic modulations.
The modulations in the potential lead to oscillations in the slow roll
parameters.
These oscillations associated with the background resonate with the
oscillations of the scalar and tensor perturbations at sub-Hubble
scales for suitable values of the parameters of the model. 
This resonance leads to persistent oscillations in the two and three-point 
functions.
The scalar and tensor power spectra as well as the scalar bi-spectrum have
been analytically evaluated earlier in the axion monodromy model under certain 
approximations.

\par

In terms of their hierarchy, after the scalar bi-spectrum, the 
scalar-scalar-tensor cross-correlation proves to be the most 
important of the three-point functions.   
In this work, we have analytically calculated the scalar-scalar-tensor
three-point function in the axion monodromy model in the same
approximation under which the scalar and tensor power spectra and the
scalar bi-spectrum had been evaluated earlier.
We find that the analytical results we have obtained match the corresponding
numerical results very well for a range of the parameters involved.
Subsequently, using the analytical results, we have also been able to 
explicitly verify the consistency relation governing the three-point 
function.

\par

The template that we have obtained here can be used to compare the 
inflationary models with the CMB data at the level of three-point
functions involving the tensor perturbations (in this context,
see the recent work, Ref.~\cite{Meerburg:2016ecv}). 
Clearly, it will be interesting to extend our analysis to the 
scalar-tensor-tensor three-point function as well as the tensor 
bi-spectrum.
We find that the tensor bi-spectrum can be easily evaluated using 
the methods adopted here. 
However, comparison with the corresponding numerical results suggest 
that these methods do not prove to be adequate to evaluate the 
scalar-tensor-tensor three-point function to the same level of accuracy.
Also, for instance, the consistency relation for the three-point function 
also does not seem to hold under these approximations.
These approximations need to be extended in order to evaluate the 
scalar-tensor-tensor cross-correlation analytically to a good level of accuracy.
We are currently investigating this issue.

%%%%%%%%%%%%%%%%%%%%%%%%%%%%%%%%%%%%%%%%%%%%%%%%%%%%%%%%%%%%%%%%%%%%%%%%%%%%%%%

\section*{Acknowledgements}

We acknowledge the use of the high performance computing facility at the
Indian Institute of Technology Madras, Chennai, India. 
DC would like to thank the Indian Institute of Technology Madras, Chennai, India, 
for financial support through half-time research assistantship.
VS would like to acknowledge support from NSF Grant No. PHY-1403943.
Portions of this research were conducted with the high performance computing resources 
provided by Louisiana State University (http://www.hpc.lsu.edu), Baton Rouge, U.S.A..
LS also wishes to thank the Indian Institute of Technology Madras, Chennai, 
India, for support through the New Faculty Seed Grant.

%%%%%%%%%%%%%%%%%%%%%%%%%%%%%%%%%%%%%%%%%%%%%%%%%%%%%%%%%%%%%%%%%%%%%%%%%%%%%%%
\bibliographystyle{JHEP}
\bibliography{amm-november-2016}

%%%%%%%%%%%%%%%%%%%%%%%%%%%%%%%%%%%%%%%%%%%%%%%%%%%%%%%%%%%%%%%%%%%%%%%%%%%%%%%

\end{document}